\documentstyle[prb,aps,twocolumn]{revtex}
\begin{document}
\title{
 \vskip -1 cm
 {\normalsize \hfill  TITCMT-95-5 }\\
 {~} \\
 Propagating S=1/2 particles in S=1 Haldane gap systems}

\author{Karlo Penc\cite{per,empenc} and Hiroyuki Shiba\cite{emshiba}
\\ {\it Department of Physics, Tokyo Institute of Technology}\\
 {\it Oh-Okayama 1-1-12, Meguro-ku, Tokyo 152, Japan }
\parbox[t]{14truecm}{\small
 Inspired by the recent experiments\cite{ditusa} on Y$_{2-x}$Ca$_x$BaNiO$_5$,
 we discuss the dispersion relation of the $s$=1/2 particles in the $s$=1
Heisenberg and VBS model in the limit of small hopping amplitudes.
The effective $s$=1/2 edge spins mix with the spin of the impurity resulting
in one four--fold and two two--fold degenarate bands.
We briefly discuss the interaction between the $s$=1/2 particles arising from
the background.
} }

\date{ February 21, 1995 }
\maketitle


Recently, the family of the quasi one--dimensional materials showing the
Haldane gap \cite{haldane} has been enlarged by the charge transfer
insulator\cite{exper} Y$_2$BaNiO$_5$. In this materials the Ni$^{2+}$ ions
with $s$=1 are surrounded by oxygens forming an octahedron,
NiO$_6$.  These octahedra are linked and form well separated chains, making
an ideal one--dimensional $s$=1 antiferromagnet. Susceptibility and neutron
scattering measurements have observed a spin gap of $\approx$ 100K (9 meV).
The two relevant Ni orbitals are $3d_{3z^2-r^2}$ and $3d_{x^2-y^2}$. The
latter is almost localized, while the $3d_{3z^2-r^2}$ orbital has finite
overlap with $2p_z$ orbital\cite{band} of the O.

 The importance of Y$_2$BaNiO$_5$ is not only being another Haldane gapped
system, but by replacing the off--chain Y$^{3+}$ by Ca$^{2+}$ one can dope
the chain by holes; thus one can introduce carriers into a gapped spin
liquid. The doped material, Y$_{2-x}$Ca$_x$BaNiO$_5$, has been studied by
DiTusa {\it et al.} (Ref.~\onlinecite{ditusa}).  The addition of carriers
lowers the resistivity and its temperature dependence cannot be described as
thermal activation across the charge gap.  Polarized X-ray absorption
experiment showed that upon doping, the holes go to the $2p_z$ orbital of the
O between the Ni$^{2+}$ ions.  Furthermore, new states with $S$ between 1 and
3/2 per impurity appeared inside the Haldane gap.

 It is not clear at the moment, what a minimal microscopic model capable to
describe the experiments can be. Neither is clear the importance of the
disorder caused by replacing Y by Ca  (the effect of the bond disorder was
addressed in Refs.~\onlinecite{sorensen,yulu}). If the disorder can be
neglected, then, as far as the low energy physics is concerned, the effective
model Hamiltonian can be that of the $s$=1 Heiesenberg model, where additional
$s$=1/2 objects are added (see Fig.~\ref{fig:nio}). The question of
localized $s$=1/2 impurities was discussed in
Refs.~\onlinecite{sorensen,nonmoving}.  However, nothing can exclude that the
holes can move, and it raises a very interesting question: what are the low
energy properties of doped Haldane systems?  For this reason here we discuss a
simple Hamiltonian to describe such systems. It turns out that even in that
simple Hamiltonian, to solve the propagation of one spin $s$=1/2 object is not
trivial, and the interaction between the spin 1/2 objects due to the $s$=1
background is rather complicated.

A similar problem is the propagation of $s$=0 holes in the Haldane gap
systems\cite{spin0}.  However, it corresponds to a simultaneous hopping of
two electrons, which can be favorable to one electron hopping only in limited
circumstances.

 To describe the interaction between the $s$=1 spins of the Ni$^{2+}$ ions,
we consider the following rotationally invariant spin Hamiltonian
\begin{equation}
  H_{0} = J
   \sum_{i}
     \left[
       {\bf S}_i {\bf S}_{i+1}
      -\beta \left({\bf S}_i {\bf S}_{i+1}\right)^2
     \right] ,
  \label{eq:Hhost}
\end{equation}
We will concentrate on the special cases of $\beta=0$ and $\beta=-1/3$. The
former is the Heisenberg model for $s$=1, while the latter is the Valence
Bond Solid (VBS) Hamiltonian, where the ground state function is
known\cite{VBS}.  Both Hamiltonians have a gap  in the excitation spectrum.
Since the holes break the valence bonds of the $s$=1 host, we get open chains
segments between the $s$=1/2 objects (see Fig.~\ref{fig:opc}).
The ground state of an open chain is
four--fold degenerate\cite{edges} in the thermodynamic limit.  This can be
simply explained by noticing that the construction of the VBS state of an
open chain leaves effective 1/2 spin objects at the ends of the chain. These
edge spins can form a triplet and a singlet, and the energy splitting for
$N$--site long chain is $\sim \exp(-N/\xi)$ and it disappears for
$N\rightarrow\infty$, where $\xi$ is the correlation length. One can actually
see these edge states, and their extension is over several lattice
sites\cite{edges2}. For $\beta=-1/3$, the correlation length is $\ln 3$,
while for the Heisenberg model $\xi \approx 7$. We denote the triplet wave
function by $T^{\pm 1,0}_N$ and singlet by $S_N$, where the superscript $\pm
1,0$ denotes the $z$ component of the triplet.  For $N$ even, the singlet is
even and triplet is odd in parity, while for $N$ odd the parities are
reversed.

 Now, let us turn to the Hamiltonian describing the effect of $s$=1/2
impurities. First of all, an impurity at site $i+1/2$ will change the
effective $J$ coupling to
$J_1$ between the $s$=1 spins at site $i$ and $i+1$ in the Hamiltonian in
Eq.~(\ref{eq:Hhost}).
Next, there is an interaction between the 1/2 impurity and $s$=1 spins
(see Refs.~\onlinecite{sorensen,nonmoving}):
\begin{equation}
 H_{J'}  = 2 J' \sum_{i}
                 \left(
                     {\bf S}_{i} \bbox{ \sigma}_{i+1/2} +
                     {\bf S}_{i+1} \bbox{ \sigma}_{i+1/2}
                 \right) \>,
  \label{eq:intnm}
\end{equation}
and finally, there are the Hamiltonians describing the effective hopping of
 the holes
on the O-sites:
\begin{eqnarray}
 H_{\tilde t}  &=&
  \tilde t \sum_{i} \hat P_{i} \nonumber\\
 H_{\tilde J} &=&
    2 \tilde J
   \sum_{i}
       \hat P_{i}
                 \left(
                     {\bf S}_i \bbox{ \sigma}_{i+1/2} +
                     \bbox{ \sigma}_{i-1/2} {\bf S}_{i}
                 \right) \>,
  \label{eq:Hhop}
\end{eqnarray}
where the factor 2 in front of the $\tilde J$ and $J'$ is for convenience
and $\bbox{ \sigma}$ is the spin operator of the $s$=1/2 spin.
The operators $\hat P_{i}$ exchanges the occupation of the site $i+1/2$ and
$i-1/2$, i.e.  if the hole is on site $i+1/2$, the $\hat P_{i}$ will move it
to $i-1/2$ if that site was empty. We do not allow twon $s$=1/2 objects to
occupy the same site, as it costs lot of energy.  Clearly, the Hamiltonian
$H = H_{0}+H_{J'}+H_{\tilde t}+H_{\tilde J}$ is isotropic.

In the following we will restrict ourselves to the case when the parameters
$\tilde t$ and $\tilde J$ describing the propagation of the $s$=1/2 objects
are small compared to the Haldane gap, i.e. $J$ of the host $s$=1 system.
Furthermore, $J_1$ and $J'$ are assumed small as well. In this limit the
energetically large excitation of the $s$=1 host
created during the motion can be neglected and we will work in the subspace
where the wave function of the $s$=1 sequences are the singlet and triplet
wave function described above. Our approach is similar to the variational
wave function applied in Ref.~\onlinecite{nonmoving}.

 Given the Hamiltonian and the constraints above, let us now turn to the
calculation of the dispersion relation of a single hole.
 First, let us construct the trial wave function of the $L$-site periodic
chain with one impurity:
\begin{eqnarray}
  |S_L\sigma k \rangle = \frac{1}{\sqrt{L}}
    \sum e^{ik j}|S_L\sigma;j\!+\!1/2 \rangle \nonumber\\
  |T_L\sigma k \rangle = \frac{1}{\sqrt{L}}
    \sum e^{ik j}|T_L\sigma;j\!+\!1/2 \rangle  \> ,
\end{eqnarray}
where $|S_L\sigma;j\!+\!1/2 \rangle$ and $|T_L\sigma;j\!+\!1/2 \rangle$
denote the states with hole at site $j\!+\!1/2$.

Now it is a good point to say something about the addition of angular momenta.
The state of the two edge 1/2 spins and the impurity 1/2 spin has one $S=3/2$
and two $S=1/2$ representations. Since our Hamiltonian is
rotationally invariant, we expect one four--fold degenerate band with total
spin $S=3/2$ (quartet) and two two--fold degenerate bands with $S=1/2$
(dublet). The $S=3/2$ representation can be constructed as follows:
\begin{eqnarray}
 |Q^{+3/2}_k\rangle &=& |T^+_L\!\uparrow k \rangle
   \nonumber\\
 |Q^{+1/2}_k\rangle &=&
    \sqrt{1/3}\> |T^+_L\!\downarrow k \rangle
  + \sqrt{2/3}\> |T^0_L\!\uparrow k \rangle
   \nonumber\\
 |Q^{-1/2}_k \rangle &=&
     \sqrt{1/3}\> |T^-_L\!\uparrow k \rangle
   + \sqrt{2/3}\> |T^0_L\!\downarrow k \rangle
   \nonumber\\
 |Q^{-3/2}_k \rangle &=& |T^-_L\!\downarrow k \rangle \>,
\end{eqnarray}
while the states belonging to the $S=1/2$ representations are
\begin{eqnarray}
 |D^{+1/2}_k \rangle &=&
    \sqrt{2/3}\> |T^+_L\!\downarrow k \rangle
  - \sqrt{1/3}\> |T^0_L\!\uparrow k \rangle
 \nonumber\\
 |D^{-1/2}_k \rangle &=&
    \sqrt{2/3}\> |T^-_L\!\uparrow k \rangle
  - \sqrt{1/3}\> |T^0_L\!\downarrow k \rangle
\end{eqnarray}
and
\begin{eqnarray}
 |\bar D^{+1/2}_k\rangle &=&  |S^0_L\!\uparrow k \rangle \nonumber\\
 |\bar D^{+1/2}_k\rangle &=&  |S^0_L\!\downarrow k \rangle \> .
\end{eqnarray}

The  parity of the states defined above is different, and is determined by
the parity of the $|T_L \rangle $ and $|S_L \rangle $. For example, if we
define the reflection operator $\hat R$ so that it changes the spin at site
$j$ to $-j$, then $\hat R |Q_k\rangle = r |Q_k\rangle$, $\hat R |D_k\rangle =
r |D_k\rangle$ and $\hat R | \bar D_k\rangle = -r |\bar D_k\rangle$, where
$r=-1$ for $L$  even and $r=+ 1$ for $L$ odd.

The rotational invariance of the Hamiltonian implies that the Hamiltonian
matrix among $|Q_k \rangle $ is
diagonal, and there are matrix elements between the $|\bar D_k\rangle$ and $|
D_k\rangle$ states only. Furthermore, since the parity of the states  $|\bar
D^{\pm 1/2}_k\rangle$ and $| D^{\pm 1/2}_k\rangle$ is different, the matrix
elements with some hermitian operator $\hat A$ commutable with $\hat R$ will
satisfy $\langle  D_k|\hat A|\bar D_k\rangle = - \langle  D_{-k}|\hat A|\bar
D_{-k}\rangle$. These arguments give
\begin{eqnarray}
  H |D_k \rangle &=& (\varepsilon_T + b_k) |D_k \rangle
            + i c_k |\bar D_k \rangle \nonumber\\
  H |\bar D_k \rangle &=&-i c_k |D_k \rangle
           +  (\varepsilon_S + d_k) |\bar D_k \rangle \nonumber\\
  H |Q_k \rangle &=& (\varepsilon_T + a_k)  |Q_k \rangle \>.
\end{eqnarray}
Here $\varepsilon_S$ and $\varepsilon_T$ are the energies of the
$|S_L \rangle $ and $|T_L \rangle$, respectively, and for small values of the
interaction parameters compared to $J$ they depend only on $J'$ apart from the
finite size effects mentioned earlier.
Then, the dispersion relation is
\begin{eqnarray}
 \varepsilon_D^{\pm}(k) &=&
    \frac{b_k\!+\!\varepsilon_T\!+\!d_k\!+\!\varepsilon_S}{2}
  \pm\sqrt{
       \left(
          \frac{ b_k \!+\! \varepsilon_T \!-\! d_k \!-\! \varepsilon_S}{2}
       \right)^2
      + c_k^2 }
  \nonumber\\
 \varepsilon_Q(k) &=& \varepsilon_T + a_k \>.
  \label{eq:disprel}
\end{eqnarray}

  We can give more precise statements about the parameters using the
microscopic model of the hopping, Eq.~(\ref{eq:Hhop}). First, let us consider
the hopping due to $H_{\tilde t}$.  Since $H_{\tilde t}$ is a scalar operator
 in the spin space, it will
have purely diagonal matrix elements:
$\langle T^m \sigma k| H_{\tilde t} |T^{m'} \sigma' k\rangle = 2\delta_{mm'}
\delta_{\sigma\sigma'}  \tilde t h^Q_t \cos k$
and
$\langle S \sigma k|  H_{\tilde t} |S \sigma' k\rangle
= 2 \delta_{\sigma\sigma'} \tilde t h^D_t \cos k$.
However, $H_{\tilde J}$ is a vector
operator in the Hilbert space spanned by the $s$=1 spins, and we can use the
Wigner--Eckart theorem to get the following identities:
$\langle T^m \sigma k| H_{\tilde J} |T^m \sigma k\rangle
= 4 m \sigma \tilde J h^Q_J \cos k$
and
$\langle T^{+1} \!\downarrow;k| H_{\tilde J}|T^0 \!\uparrow;k\rangle
= 2 \sqrt{2} \tilde J h^Q_J \cos k$,
furthermore $\langle S \sigma;k| H_{\tilde J} |S \sigma';k\rangle = 0$.
Similarly,
$\langle T^{+1} \!\downarrow;k| H_{\tilde J} |S \!\uparrow;k\rangle
= - i 2 \sqrt{2}\tilde J h^D_J \sin k$ and
$\langle T^{0} \!\uparrow;k| H_{\tilde J} |S \!\uparrow;k\rangle
=  i 2 \tilde J h^D_J \sin k$.
The same arguments  can be applied for the matrix elements of $H_{J'}$ as
well. Putting all together, we get
\begin{eqnarray}
  a_k  &=& 2 (h^Q_t \tilde t + h^Q_J \tilde J) \cos k + J' g \nonumber\\
  b_k  &=& 2 (h^Q_t \tilde t -2 h^Q_J \tilde J) \cos k -2 J' g \nonumber\\
  c_k  &=& 2 \sqrt{3} h^D_J \tilde J \sin k \nonumber\\
  d_k  &=&  2 h^D_t \tilde t  \cos k \label{eq:hm1}
\end{eqnarray}
where the coefficients $h^Q_t, h^Q_J, h^D_t$, $h^D_J$ and $g$ depends on the
wave functions $|S_L\rangle$ and $|T_L \rangle$ and their size dependence is
governed by the $\xi$.

Here we will calculate
these coefficients for the VBS model and the Heisenberg model.

 {\it VBS model:} For the VBS model the task is essentially simplified due
to the knowledge of
how to construct the ground state wave function. Actually, for our purposes,
the following decomposition of the four lowest lying wave functions turns out
to be useful:
\begin{eqnarray}
    | T^+_N \rangle
      & = & \alpha_N  |T^+_{N-1} 0\rangle
        -   \alpha_N  |T^0_{N-1}\!+\!\rangle
        -   \beta_N   |S_{N-1}\!+\!\rangle \nonumber\\
    | T^0_N \rangle
      & = & \alpha_N  |T^+_{N-1}\!-\!\rangle
        -   \alpha_N  |T^-_{N-1}\!+\!\rangle
        -   \beta_N  |S_{N-1}0\rangle \nonumber\\
    | T^-_N \rangle
      & = & \alpha_N  |T^0_{N-1}\!-\!\rangle
        -   \alpha_N  |T^-_{N-1}0\rangle
        -   \beta_N  |S_{N-1}\!-\!\rangle \nonumber\\
    | S_N \rangle
      & = & \sqrt{1/3} \> \left(  |T^+_{N-1}\!-\!\rangle
        -     |T^0_{N-1}0\rangle
        +     |T^-_{N-1} \!+\!\rangle \right) \nonumber
\end{eqnarray}
and similarly
\begin{eqnarray}
    | T^+_N \rangle  & = &
       -  \alpha_N |0 T^+_{N-1}\rangle
       + \alpha_N  |\!+\! T^0_{N-1}\rangle
       -  \beta_N  |\!+\!S_{N-1}\rangle \nonumber\\
    | T^0_N \rangle   & = &
       -  \alpha_N  |\!-\!T^+_{N-1}\rangle
       + \alpha_N |\!+\!T^-_{N-1}\rangle
       -  \beta_N |0 S_{N-1}\rangle \nonumber\\
    | T^-_N\rangle & = &
       - \alpha_N  |\!-\! T^0_{N-1}\rangle
       + \alpha_N |0T^-_{N-1}\rangle
       -  \beta_N  |\!-\!S_{N-1}\rangle \nonumber\\
    | S_N \rangle & = & \sqrt{1/3} \> \left(
          |\!-\! T^+_{N-1}\rangle
       -  |0 T^0_{N-1}\rangle
       +  |\!+\! T^-_{N-1}\rangle\right) \nonumber   \>,
\end{eqnarray}
where the coefficients $\alpha_N$ and $\beta_N$ are given in Tab.~\ref{tab:1}.
It is easy to get this decomposition by inspection. We believe that it is
trivial to get this result from the transfer--matrix\cite{zittartz}
representation of the VBS wave function.  The coefficients exhibit the
following remarkable properties: $2 \alpha_N^2 + \beta_N^2 =1$, and it means
that in this decomposition is complete in the Hilbert space spanned by the
four
VBS wave functions. Furthermore, $\alpha_{N} = (2+3\alpha^2_{N-1})^{-1/2} $,
which allows us to calculate the coefficients recursively.

  With the help of the wave function decomposition presented above, it is
straightforward to get the parameters for the hopping matrix elements:
\begin{eqnarray}
   h^Q_t &=& \beta_L^2 - 2 \alpha^2_L  \nonumber\\
   h^Q_J &=& \beta_L^2 - \alpha^2_L  \nonumber\\
   h^D_t &=& 1 \nonumber\\
   h^D_J &=& 2 \alpha_L / \sqrt{3} \nonumber\\
   g &=& 2 (\alpha_L^2 + \beta_L^2)
\end{eqnarray}
furthermore $\varepsilon_S = -2 J_1/3$ and
$\varepsilon_T = (34 - 80 \alpha_L^2) J_1/9$, where the energy is measured
from $\varepsilon_T(J_1=0) = \varepsilon_S(J_1=0)$.
Let us comment here that $h^D_t = 1$ means that no walls were created during
the motion for that special process.

For infinitely large system and $J_1=0$ the dispersion relation
Eq.~(\ref{eq:disprel}) is
simplified to
\begin{eqnarray}
  \varepsilon_D^{\pm}(k) & = & \frac{2}{3} \tilde t \cos k
   - \frac{4}{3} J'
   \pm
   \frac{4}{3} \sqrt{
     3 \tilde J^2 \sin^2 k
     + (\tilde t \cos k + J')^2
    } \nonumber \\
  \varepsilon_Q(k) &= & -\frac{2}{3} \tilde t \cos k + \frac{4}{3} J' \>.
\end{eqnarray}
We show some examples of the dispersion relation in Fig.~\ref{fig:disprel}.
  An interesting feature of the dispersion relation is that for
large values of $\tilde J$
the minimum moves away from $k=\pi$ (when $\tilde t>0$) or $k=0$ (if
 $\tilde t<0$). It means that the holes will be described by a two--band
model, which can have interesting features.

{\it Heisenberg model:} We also calculated the hopping matrix elements in
 Eq.~(\ref{eq:hm1}) for small
clusters of up to 15 sites for the more realistic Heisenberg model, where
$\beta =0$.  In that case the correlation length is comparable to the
cluster size and the size dependence of the matrix elements is large.
We have plotted the different matrix element on Fig.~(\ref{fig:matele}).
Although
the size is not large enough to get good values for $L \rightarrow \infty $
limit, we can conclude that $h^Q_t =-0.28\pm 0.01 $, $h^Q_J = 0\pm 0.005$,
$h^D_t = 0.81 \pm 0.01$ and $h^D_J =0.55\pm 0.01$. Furthermore, from
Ref.~\onlinecite{sorensen} we know that $g = \alpha$ and
$\varepsilon_S-\varepsilon_T \approx \alpha^2 J_1$ for $J_1 \ll J$, where
$\alpha = 1.0640$. We find these matrix elements to be $\approx$ 20 \% less
then those of the VBS Hamiltonian.

 A few words about the validity of the approach presented above. During the
motion the spin 1/2 object can destroy the hidden AF order\cite{walls} by
creating walls. Taking this into account, it would give us corrections of the
order $\tilde t^2/J$ and $\tilde J^2/J$ to the dispersion relation. Also with
the increasing amplitude of the hopping, the upper bands will merge with the
continuum of the states above the gap. On the other hand, the numerical
calculation on small clusters shows that the qualitative features of the
lowest band remains even when the hopping amplitudes are comparable with the
magnitude of the Haldane gap.

Now, let us turn to the question of what happens if there
are more than one $s$=1/2 impurities? We can follow the idea that for small
values of the $\tilde t$ and $\tilde J$, the states above the gap are not
excited and it is enough to consider the four low-lying states of the open
chain for the wave--function of the $s$=1 sequence in the wave functions. For
example, a typical state is
\begin{equation}
    | \dots  T^+_{i_1-i_0} \uparrow_{i_1} T^0_{i_2-i_1} \downarrow_{i_2}
    S_{i_3-i_2}
   \downarrow_{i_3}  \dots \rangle  \label{eq:varwf}.
  \end{equation}
We can think of this wave function as an variational ansatz. Then, the
interaction between the $s$=1/2 objects in this wave function comes from:
(i) the hopping matrix elements, which depends strongly on the size of the
open
chains, i.e. on the distance of the nearest holes; (ii) the energy splitting
of the singlet and triplet states of the finite chains between the holes.
 The interaction due to (i) is proportional to the hopping
amplitudes $\tilde t$ and $\tilde J$ itself, while the strength of (ii) is
determined by the $J$ of the host system.
These effects depend very much on the correlation length of the $s$=1 system,
and they are the smallest for the VBS model. Actually in that case the energy
splitting is zero and the interaction is due to (i) only. These interactions
can in principle result in a bound state, unless the kinetic energy is large
enough.

 Despite of the strong constraints involved in construction of the
variational wave function Eq.~(\ref{eq:varwf}), it has still a substantial
freedom and the properties of the system with more impurities remains
to be solved.


As far as the experiments done on Y$_{2-x}$Ca$_x$BaNiO$_5$ are concerned, we
have
shown that there are states with $S$ larger than the $s$=1/2 of the impurity
in the Haldane gap. Unfortunately, the parameter range for the
Y$_{2-x}$Ca$_x$BaNiO$_5$ based on simple electronic model seems to indicate
that
the hopping amplitudes are comparable with the interaction between the $s$=1
spins, where our approach is valid only qualitatively.

\begin{table}
\caption{The coefficients of the VBS wave function decomposition\label{tab:1}}
\begin{tabular}{rccccccc}
 $N$   & 1 & 2 & 3 & 4 & 5 & $\dots$ & $\infty$ \\
 $\alpha_N$ & 0 & $\sqrt{1/2}$ &  $\sqrt{2/7}$ &  $\sqrt{7/20}$ &
    $\sqrt{20/61}$ & $\dots$ & $\sqrt{1/3}$ \\
 $\beta_N$ & 1 & 0  &  $\sqrt{3/7}$  &  $\sqrt{6/20}$
  &  $\sqrt{21/61}$  & $\dots$ &  $\sqrt{1/3}$  \\
\end{tabular}
\end{table}

\begin{figure}
\caption{Schematic diagram of the Ni--O chain with typical low--energy level
  occupation. The hole is on the second O.
 \label{fig:nio}}
\end{figure}

\begin{figure}
\caption{ Part of the $s$=1 chain with two $s$=1/2 spins. Between them the
three $s$=1 spins form an open chain.
 \label{fig:opc}}
\end{figure}

\begin{figure}
 \caption{Dispersion relation for different values of $\tilde J/\tilde t$
  and $J'=J_1=0$.
  The dashed line is $\varepsilon_Q(k)$, the lower solid line is for
  $\varepsilon_D^-(k)$ and the upper for $\varepsilon_D^+(k)$. Energy is
measured from $\varepsilon_S=\varepsilon_T$.
 \label{fig:disprel}}
\end{figure}

\begin{figure}
\caption{The matrix elements $h$ for the Heisenberg model:
  $h^D_t$ (triangles),
  $h^D_J$ (diamonds),
  $h^Q_J$ (squares) and
  $h^Q_t$ (hexagons) from the top to the bottom.
\label{fig:matele}}
\end{figure}

\end{document}